\documentclass[aps,prd,preprint,groupedaddress,showpacs,showkeys]{revtex4}
\usepackage{amssymb,bm}
\usepackage[dvips]{graphicx}

\begin{document}
\bibliographystyle{apsrev}

\title{Decays of Fourth Generation Bound States}
\author{V.F. Dmitriev }
\email{V.F.Dmitriev@inp.nsk.su}
\affiliation{Budker Institute
of Nuclear Physics, 630090 Novosibirsk, Russia \\
and\\
Novosibirsk State University}
\author{V.V. Flambaum}
\email{Flambaum@phys.unsw.edu.au}
\affiliation{School of Physics, University of New South Wales, Australia}

\date{\today}
\begin{abstract}
We consider the decay modes of the heavy $q^\prime\bar{q^\prime}$ bound states originating from Higgs boson exchange between quark -- anti-quark pair. In case of a
small coupling between the fourth and lower generation the main decay mode is $q^\prime\bar{q^\prime}$ annihilation. We show that for a vector state the dominant
decay modes are Higgs-gamma and Higgs-Z decays, while for a pseudoscalar state the strong two-gluon decay mode dominates. The bound states are very narrow. The
ratio of the total width to the binding energy is less than 1\% if we are not extremely close to  the critical quark mass where the binding energy is very small.
The discussed decay modes exist for any fermion-antifermion bound states including heavy leptons and heavy neutrinos if their masses are high enough to form a bound state due to attractive Higgs boson exchange potential.
\end{abstract}
\pacs{14.40.Pq,13.25.Jx }
\keywords{Higgs - gamma, Higgs - Z, decays   }
\maketitle
\section{Introduction}
It is generally claimed that the Standard Model (SM) comprises three generations of
fermions. However, many fundamental problems do not find an answer in this framework,
and a possibility of additional massive fermions, such as a new sequential family of quarks,
is currently among the models in the spotlight of experimental searches at the LHC (for
an overview, see, e.g., \cite{hh09}). Besides its phenomenological relevance, the Standard Model
with a fourth generation  can also serve as a template for new physics
models for which the unitarity of the $3\times 3$ CKM matrix might be violated \cite{af07}. Despite
the absence of hints in experimental searches, there is currently a renewed interest in the
fact that new fermion families are still allowed by the electroweak precision constraints if
additional parameters are considered \cite{bgm12}.

The strength of the attraction between heavy
particles due to the Higgs boson exchange increases with
the particle mass. This may lead to formation of a new
type of bound states \cite{iw11,fk11}. The Higgs-induced bags made of
heavy fermions and bosons have been considered in numerous
publications - see, e.g., [ 6 \text{-} 30 ]. In this paper we shall discuss some exotic decay modes of the simplest bound states
 - $q\bar{q}$-color singlet states, although these decay modes exist for bound states of heavy leptons as well.  The decay of color-octet states was discussed in ref. \cite{ehy11}.

\section{Vector States}
Since the mass difference between $t^\prime$ and $b^\prime$ quarks is expected to be less than $W$ mass \cite{kribs07}, and the CKM matrix elements between the fourth and lower generations
could be small \cite{bgm12}, the main decay channels are the annihilation ones.
\subsection{Strong Decays}
A strong decay width for $S=1$ states is given in lowest order in $\alpha_s$ by the well known three-gluon width:
\begin{equation}\label{1}
\Gamma_{3g}=\frac{40}{81}(\pi^2-9)\frac{\alpha_s^2}{m_f^2}|\Psi(0)|^2.
\end{equation}
 Here $\alpha_s$ is the strong QCD constant at the scale $\sim m_f$ and $\Psi(0)$ is the wave
 function of $q\bar{q}$-bound state at the origin.
\subsubsection{Higgs Decays}
Among possible decay modes of the bound states of $t^{\prime}
\bar{t}^{\prime}$ or $b^{\prime} \bar{b}^{\prime}$ quarks there could be an
exotic decay mode to Higgs bosons.
It is worth  noting that decay to any number of Higgs bosons is absent for a
pseudoscalar state. (It is impossible to construct a pseudoscalar symmetric
under permutation of bosons momenta.) For vector states the minimal number of
emitted scalar bosons could be three. Two identical scalar particles cannot be in a
state with orbital angular momentum equal to 1. However, for s-states total spin of $q\bar{q}$ pair is essentially decoupled from orbital motion and for the scalar Higgs boson there is no difference between $S=0$ and $S=1$ states. Therefore, the 3H-decay is suppressed for the s-states. For p-states, due to spin-orbit coupling, the decay in three Higgs bosons is allowed.

For s-states with $S=1$ there is, however, a possibility of combined decay to Higgs boson + $\gamma$, or
Higgs boson + $Z$ boson. The amplitude for $H\gamma$ or $HZ$  channels can be presented by the equation:
\begin{equation} \label{1.1}
 T_{ch}(\bm{k})=\int\frac{d^3p}{(2\pi)^3}M_{ch}(\bm{p},\bm{k})\Phi(\bf{p}),
\end{equation}
where $ch$ stands for $H\gamma$ or $HZ$, $M_{ch}(\bm{p},\bm{k})$ is the annihilation amplitude of a free quark-antiquark pair with the relative momentum $\bm{p}$ and the momentum $\bm{k}$ of
$\gamma$-quantum or $Z$ boson. $\Phi(\bm{p})$ is a Fourier component of the spatial wave function of a  $q\bar{q}$-pair in the quarkonium. For a non-relativistic system
\begin{equation} \label{1.2}
 T_{ch}(\bm{k})= M_{ch}(0,\bm{k})\Psi(0),
\end{equation}
where $\Psi(0)=\int\frac{d^3p}{(2\pi)^3}\Phi(\bf{p})$ is the spatial wave function at the origin, $r=0$. For $H\gamma$ decay the amplitude $M_{H\gamma}(\bm{p},\bm{k})$ is equal to
\begin{equation} \label{1.3}
M_{H\gamma}(\bm{p},\bm{k})=(\sqrt{2}G_Fm_f^2)^{1/2}e_q\bar{u}(-p^\prime)\left(\frac{1}{\hat{p}-\hat{k}-m_f}\gamma_\mu+\gamma_\mu\frac{1}{-\hat{p^\prime}+\hat{k}-m_f}\right)u(p)\epsilon^\mu_\lambda(k).
\end{equation}
Here $G_F$ is the Fermi constant, $m_f$ is the quark mass, and $e_q$ is the quark electric charge. $p$ and $p^\prime$ are the 4-momenta of the quark and anti-quark, $k$ is the 4-momentum
of the emitted $\gamma$-quantum, and $\epsilon^\mu_\lambda(k)$ is the polarization vector of the  $\gamma$-quantum with the polarization $\lambda$. At $\bm{p}=0$ the amplitude simplifies considerably. Dirac 4-spinors become effectively two component ones, $u(p)=\left(\begin{array}{c}
                                                                      \phi \\
                                                                      0
                                                                    \end{array}\right)$, and the amplitude  becomes
 \begin{equation} \label{1.4}
M_{H\gamma}(0,\bm{k})=\frac{(\sqrt{2}G_Fm_f^2)^{1/2}e_q}{M}2\sqrt{2}\bm{e}_\sigma\cdot\bm{\epsilon}_\lambda(\bm{k}),
 \end{equation}
where we introduced the polarization vector of $q\bar{q}$ bound state $\sqrt{2}\bm{e}_\sigma=\chi^\dagger_{-\sigma_2}\bm{\sigma}\phi_{\sigma_1}$, $M$ is the mass of $q\bar{q}$ state,
$\sigma_1$ and $\sigma_2$ are the spin projections of the quark and the anti-quark.

 With this amplitude for the $H\gamma$-decay width we obtain
\begin{equation}\label{2}
\Gamma_{H\gamma}=\sqrt{2}\frac{2}{3}e_q^2\alpha G_F(1-\frac{m_H^2}{M^2})|\Psi(0)|^2,
\end{equation}
where  $\alpha$ is a fine structure constant, and $m_H$ is the Higgs mass.

Both widths, Eq.(\ref{1}) and Eq(\ref{2}), are proportional to $|\Psi(0)|^2$, therefore their ratio does not depend on details of the wave function and can be used to estimate the importance of the $H\gamma$-decay mode. In Fig.(1) we plot the ratio $R_\gamma=\Gamma_{H\gamma}/\Gamma_{ggg}$ for $t^\prime\bar{t^\prime}$ quarkonium as a function of fermion mass for the Higgs mass $m_H=125.3$ GeV within an interval of the fermion mass $500 - 1000$ GeV.   Obviously, the $H\gamma$-decay mode dominates in the whole mass interval.
\begin{figure}
\includegraphics[scale=1]{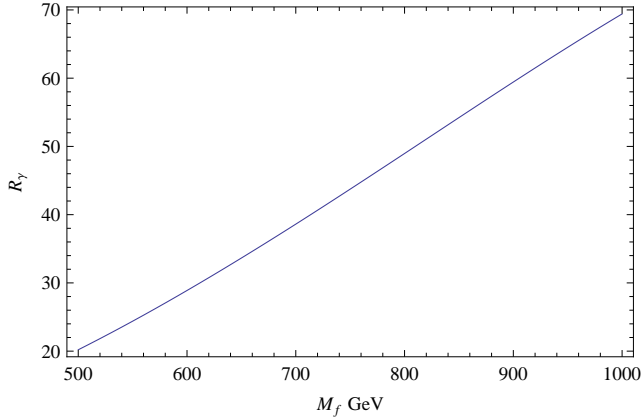}
\caption{Ratio of Higgs-$\gamma$ decay width to 3g-width as a function of quark mass.}
\end{figure}
The ratio in Fig.(1) was calculated for $t^\prime$ quarkonium. For $b^\prime$ quarkonium the ratio is smaller by a factor of 4 according to the electric charge of $b^\prime$.

The absolute value of the $H\gamma$-width for $t^\prime$ quarkonium is plotted in Fig.(2). In calculations we used the hydrogen type trial wave function from \cite{fk11}.
\begin{equation}\label{3}
\Psi(r)=\frac{1}{\sqrt{\pi}}q^{3/2}\exp(-qr),
\end{equation}
with the parameter $q$ found by minimization of the binding energy.
The binding energy of the system can be written as
\begin{equation}
 E = \langle T \rangle  + \langle V_{\text{rel}} \rangle +  \langle V_{\text{rad}} \rangle + \langle V_{\text{H}} \rangle  +  \langle V_{\text{Strong}} \rangle,
\label{octetenergy}
\end{equation}
where $ \langle V_{\text{rel}} \rangle$ and  $\langle V_{\text{rad}}\rangle$ represents relativistic and radiative corrections, respectively.
The exact expressions of all the operators in (\ref{octetenergy}) can be found in \cite{fk11}. We will present the results for their values averaged over the wave function (\ref{3}):
\begin{eqnarray}
&\frac{\langle T \rangle}{m_f} =\frac{4}{\pi}\left\{\frac{x(3-4x^2+4x^4}{3(1-x^2)^2}+\frac{(1-2x^2)\arccos(x)}{(1-x^2)^{5/2}}\right\}\\
&\langle V_{\text{H}} \rangle =-\frac{4\alpha_H q^3}{(m_H+2q)^2}\\
&\langle V_{\text{Strong}} \rangle =-\frac{4}{3}\alpha_s q\\
&\langle V_{\text{rel}} \rangle =\frac{6\alpha_H q^4}{m_f^2 (2q+m_H)}\\
&\langle V_{\text{rad}} \rangle =\frac{4\alpha_H^2q^3}{\pi m_f^2}\left(\gamma - \frac{\nu}{20}-\frac{4\pi\gamma m_H^2}{(m_H+2q)^2}\right),
\end{eqnarray}
where $x=q/m_f$, $\alpha_H=\sqrt{2}G_Fm_f^2/4\pi$,   $\nu=11$ for 4G quark - the number of heavy fermions in the polarization loop and coefficient $\gamma$ is given by the following expression:
\begin{equation}
\gamma=\frac{1}{3}\left(\ln\frac{m_f+m_H}{m_H}-\frac{7m_f}{4m_f+5m_H}\right).
\end{equation}
Minimizing the binding energy Eq.(\ref{octetenergy}) with respect to parameter $q$ in the trial function Eq.(\ref{3}) we obtain both the variational binding energy and the variational wave function used in Eqs.(\ref{1},\ref{2}). The simple non-relativistic approach can be used for quark masses $m_f \leq $1000 GeV. For heavier masses relativistic effects become significant.
\begin{figure}
\includegraphics[scale=1]{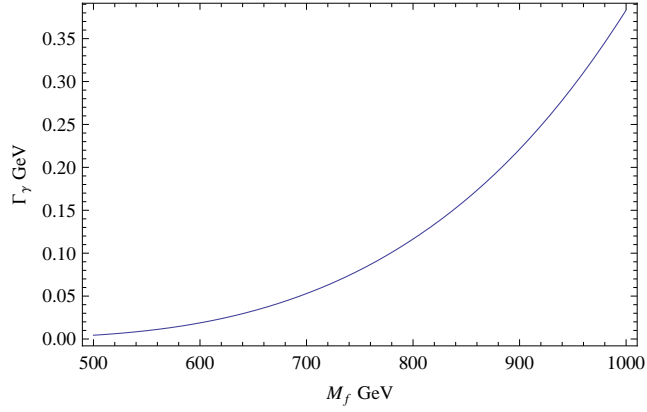}
\caption{The width $H\gamma$-decay mode as a function of the fermion mass. }
\end{figure}

The amplitude of the HZ-annihilation of free quarks is given by the equation:
$$
M_{HZ}(\bm{p},\bm{k})=\frac{(\sqrt{2}G_F m_f^2)^{1/2}e}{\sin\theta_W \cos\theta_W}\bar{u}(-p^\prime)\left[\frac{1}{\hat{p}-\hat{k}-m_f}\left(\gamma_\mu(\frac{1}{4}-\frac{2}{3}\sin^2\theta_W) + \frac{1}{4}\gamma_\mu\gamma_5\right)\right.
$$
\begin{equation} \label{4}
\left. + \left(\gamma_\mu(\frac{1}{4}-\frac{2}{3}\sin^2\theta_W) + \frac{1}{4}\gamma_\mu\gamma_5\right)\frac{1}{-\hat{p^\prime}+\hat{k}-m_f}\right]u(p)\epsilon^\mu_\lambda(k),
\end{equation}
where $\theta_W$ is the Weinberg angle. At $\bm{p}=0$ the axial current does not contribute and the amplitude Eq.(\ref{4}) differs from Eq.(\ref{1.4}) just by a factor and kinematics of massive Z boson
\begin{equation} \label{5}
M_{HZ}(0,\bm{k})= \frac{e(\sqrt{2}G_F m_f^2)^{1/2}}{m_Z^2-ME_k}\frac{\frac{1}{4}-\frac{2}{3}\sin^2\theta_W}{\sin\theta_W \cos\theta_W}2\sqrt{2}E_k\left(\bm{e}_\sigma\cdot\bm{\epsilon}_\lambda(\bm{k}) -\frac{(\bm{k}\cdot\bm{e}_\sigma)(\bm{k}\cdot \bm{\epsilon}_\lambda(\bm{k})}{E_k^2}\right).
\end{equation}
Here $E_k$ is the Z-boson energy. With this amplitude
the Higgs-Z decay width for $t^\prime\bar{t^\prime}$ quarkonium is of the following form:
\begin{equation}
\Gamma_{HZ}=\frac{8\alpha\sqrt{2}G_F m_f^2}{\sin^2\theta_W\cos^2\theta_W(m_Z^2-M E_k)^2}(\frac{1}{4}-\frac{2}{3}\sin^2\theta_W)^2(E_k^2-\bm{k}^2/3)\frac{k}{M}|\Psi(0)|^2.
\end{equation}

The ratio $R_Z=\frac{\Gamma_{HZ}}{\Gamma_{3g}}$ as a function of fermion mass is plotted in Fig.(3). The Higgs-Z width is about one order of magnitude smaller than the Higgs-gamma width.
\begin{figure}
\includegraphics[scale=1]{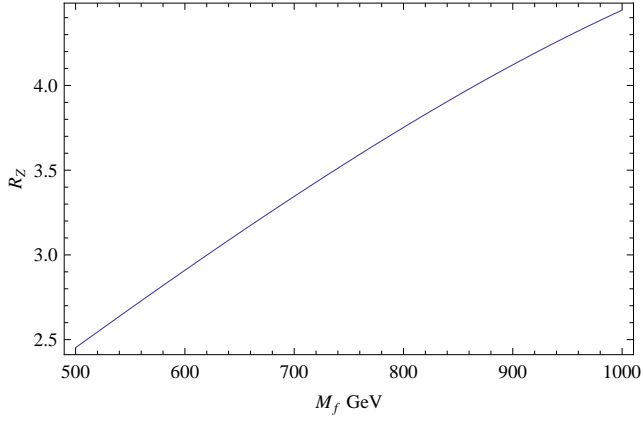}
\caption{Ratio of Higgs-Z decay width and 3g-width as a function of the fermion mass for $t^\prime\bar{t^\prime}$ state.}
\end{figure}
The suppression comes from the factor $(\frac{1}{4}-\frac{2}{3}\sin^2\theta_W)^2$ which is equal to 0.009. For $b^\prime\bar{b^\prime}$-bound state the corresponding factor is 3.25 times larger and the HZ-width becomes comparable with H$\gamma$-width.

The ratio of the total width to the $q^\prime\bar{q}^\prime$ binding energy for the vector state is shown in Fig.(4). In the whole mass interval the ratio is smaller than 1\%, therefore, in case of small mixing with lower generations, the vector bound states are long lived.
\begin{figure}
\includegraphics[scale=1]{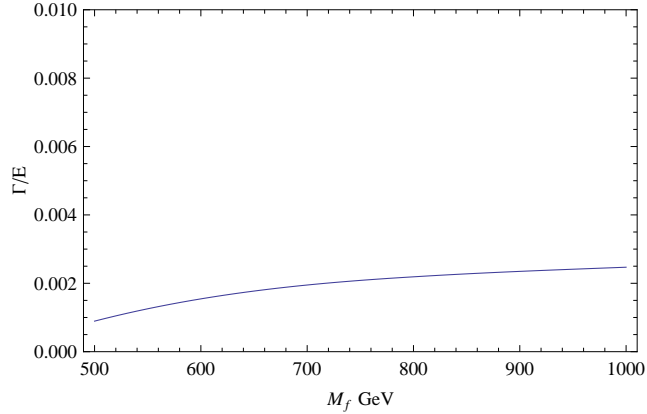}
\caption{Ratio $\Gamma_{tot}/E$ as a function of the fermion mass for $S=1$ state.}
\end{figure}

\section{Pseudoscalar States}
For pseudoscalar states the main decay mode is the two-gluon one. The decay into two gammas or two  W-bosons is much smaller.
The width for decay in two W-boson is
\begin{equation}
\Gamma_{2W}=\frac{\pi \alpha^2}{4 \sin^2\theta_W}\frac{(m_f^2-m_W^2)^{3/2}}{m_f(2 m_f^2-m_W^2)^2}|\Psi(0)|^2.
\end{equation}
At all fermion masses in the discussed range $\Gamma_{2W} \approx 0.003\Gamma_{gg}$. Therefore, the total annihilation width is small compared to the binding energy and $^1S_0$ states are also long lived in case of small mixing.
\section{conclusions}
We discussed new decay modes for bound states of  heavy $q^\prime\bar{q^\prime}$-pair. In the absence of strong mixing with lower generations, annihilation to Higgs-$\gamma$ and Higgs-Z dominates the annihilation into hadrons for $^3S_1$ states. Detection of a single high energy $\gamma$-quantum can be a clear signal of creation the heavy bound $q^\prime\bar{q^\prime}$ pair.  For $^1S_0$ state, decays to Higgs boson in final states are suppressed. The allowed decay is to W-pair. However, its width is small compared to annihilation into hadrons. In both cases the total annihilation width is small compared to binding energy, therefore the bound states are long lived. It is worth to say that the discussed decay modes exist for any fermion-antifermion bound states including heavy leptons and heavy neutrinos if their masses are high enough to form a bound state due to attractive Higgs boson exchange potential.
\acknowledgements
This work was supported in part  (VFD) by the Ministry of Science and Education of the Russian Federation, the RFBR grant 12-02-91341-NNIO\_a, and by Australian Research Council. One of the authors (VFD) acknowledges support from UNSW Faculty of Science Visiting Research Fellowships and Godfrey fund.


\begin{thebibliography}{99}

\bibitem{hh09} B. Holdom, W. S. Hou, T. Hurth, M. L. Mangano, S. Sultansoy and G. Unel, PMC Phys. A
3, 4 (2009).
\bibitem{af07}J. Alwall, R. Frederix, J.-M. Gerard, A. Giammanco, M. Herquet, S. Kalinin, E. Kou, V.
Lemaitre, F. Maltoni, Eur. Phys. J. C49, 791-801 (2007).
\bibitem{bgm12} M. Buchkremer, J.-M. Gerard, F. Maltoni, JHEP 1206, 135 (2012).
\bibitem{iw11} K. Ishiwata, M. B.Wise, Phys. Rev. D83, 074015 (2011).
\bibitem{fk11} V.V. Flambaum, M.Yu. Kuchiev, Phys. Rev. D84, 114024 (2011).
\bibitem{kfs10}M. Y. Kuchiev, V. V. Flambaum, and E. Shuryak, Phys.
Lett. B 693, 485 (2010).
\bibitem{ku10} M. Yu. Kuchiev, Phys.Rev. D82, 127701 (2010).
\bibitem{v72} P. Vinciarelli, Lett. Nuovo Cim. 4S2, 905 (1972).
\bibitem{lw74} T. D. Lee and G. C. Wick, Phys. Rev. D 9, 2291 (1974).
\bibitem{7} A. Chodos et al, Phys. Rev. D9, 3471 (1974).
\bibitem{8} M. Creutz, Phys. Rev. D10, 1749 (1974).
\bibitem{9} W. A. Bardeen et al, Phys. Rev. D11, 1094 (1975).
\bibitem{10} R. Giles, S. H. H. Tye, Phys. Rev. D13, 1690 (1976).
\bibitem{11} K. Huang, D. R. Stump, Phys. Rev. D14, 223 (1976).
\bibitem{12} R. Friedberg, T. D. Lee, Phys. Rev. D15, 1694 (1977).
\bibitem{13} R. Goldflam, L. Wilets, Phys. Rev. D25, 1951 (1982).
\bibitem{14} R. MacKenzie, F. Wilczek, and A. Zee, Phys. Rev. Lett.
53, 2203 (1984).
\bibitem{15} L. R. Dodd, M. A. Lohe, Phys. Rev. D32, 1816 (1985).
\bibitem{16} S. Y. Khlebnikov, M. E. Shaposhnikov, Phys. Lett.
B180, 93 (1986).
\bibitem{17} G. W. Anderson, L. J. Hall, and S. D. H. Hsu, Phys.
Lett. B249, 505 (1990).
\bibitem{18} R. MacKenzie, Mod. Phys. Lett. A7, 293 (1992).
\bibitem{19} A. L. Macpherson, B. A. Campbell, Phys. Lett. B306,
379 (1993).
\bibitem{20} R. Johnson, J. Schechter, Phys. Rev. D36, 1484 (1987).
\bibitem{21} C. D. Froggatt, H. B. Nielsen, Phys. Rev. D80, 034033
(2009); C. D. Froggatt et al, arXiv:0804.4506 [hep-ph], (2008).
\bibitem{22} M. Y. Kuchiev, V. V. Flambaum, E. Shuryak, Phys. Rev.
D78, 077502 (2008).
\bibitem{23} J.-M. Richard, Few Body Syst. 45, 65 (2009).
\bibitem{24} M. P. Crichigno, E. Shuryak, arXiv:0909.5629 [hep-ph], (2009).
\bibitem{25} M. P. Crichigno, V. V. Flambaum, M. Y. Kuchiev,
E. Shuryak, Phys. Rev. D 82, 073018 (2010).
\bibitem{26} M. Y. Kuchiev, and V. V. Flambaum, EuroPhys Lett. 97, 51001 (2012).
\bibitem{27} M. Yu. Kuchiev, Phys. Rev. D82, 127701 (2010).
\bibitem{ehy11} Tsedenbaljir Enkhbat, Wei-Shu Hou, and Hiroshi Yokoya, Phys. Rev. D84, 094013 (2011).
\bibitem{kribs07} G. D. Kribs, T. Plehn, M. Spannowsky, T. M. P. Tait, Phys. Rev. D76, 075016 (2007).
\end{thebibliography}
\end{document}